# Register Allocation By Model Transformer Semantics

by Yin Wang, R. Kent Dybvig

Indiana University

**Abstract**

Register allocation has long been formulated as a graph coloring problem, coloring the conflict graph with physical registers. Such a formulation does not fully capture the goal of the allocation, which is to minimize the traffic between registers and memory. Linear scan has been proposed as an alternative to graph coloring, but in essence, it can be viewed as a greedy algorithm for graph coloring: coloring the vertices not in the order of their degrees, but in the order of their occurence in the program. Thus it suffers from almost the same constraints as graph coloring. In this article, I propose a new method of register allocation based on the ideas of *model transformer semantics* (MTS) and *static cache replacement* (SCR). Model transformer semantics captures the semantics of registers and the stack. Static cache replacement relaxes the assumptions made by graph coloring and linear scan, aiming directly at reducing register-memory traffic. The method explores a much larger solution space than that of graph coloring and linear scan, thus providing more opportunities of optimization. It seamlessly performs live range splitting, an optimization found in extensions to graph coloring and linear scan. Also, it simplifies the compiler, and its semantics-based approach provides possibilities of simplifying the formal verification of compilers.

## 1 Introduction

Register allocation is a central problem of compiler construction and a profitable optimization in a compiler. The quality of the code generated by register allocation has a huge impact on the runtime performance. Thus a major part of compiler research has been devoted to this problem. Since Chaitin's seminal paper [2], register allocation has been identified with graph coloring [3], a famous NP-Complete problem. Since then, researches have been mainly to improve and extend graph coloring algorithms, without questioning the assumptions it makes. This trend has only recently been challenged by linear scan [5], which is supposedly more efficient than graph coloring in compilation speed, and thus attracted much attention because of the rise of JIT (just-in-time) compilers. However, the original linear scan algorithm [5] generates slower code than graph coloring. Only with extensions such as live range splitting [7] can linear scan outperform the vanilla graph coloring algorithm. Linear scan is not very different from graph coloring. In fact, it can be viewed as a particular graph coloring algorithm. Instead of looking at the vertices in the order of their degrees as in Chaitin's heauristics, linear scan looks at the vertices in the order of their occurrences in the input program. This explains why linear scan (without extensions) does not outperform graph coloring.

In this article, I propose a method for register allocation that sets itself apart from graph coloring and linear scan. The goal is to provide another practical method and at the same time relate the major register allocation methods together in a principled

way—explaining and predicting their advantages and drawbacks not by benchmarks, but by reasoning. The method is based on the so-called *model transformer semantics* (MTS) and *static cache replacement* (SCR). The two aspects of the method are indispensible with respect to each other and work in a seamless fashion. For convenience, in the rest of this article, we will call these two aspects of the method collectively "the MTS method" or just "MTS".

In short, a *model* is an abstract representation of the physical machine. For example, a model may specify that at a certain program point the variable $x$ must be located in register $r2$. A *model transformer* is a rule that specifies how the models relate to each other. For example, a model transformer may specify that after the instruction $x := y + 1$, the variable $y$, wherever it had been located, will be loaded into some register that is not occupied by any other variable. A *model transformer semantics* is thus a set of model transformers dispatched by the syntax of the intermediate language. In essence, it is an abstract interpreter of the intermediate language. It is also a special kind of logic (similar to Hoare Logic and linear logic), but instead of having a syntax and inference rules, it encodes the logic formulas directly into data structures (models) which *entail* the static properties of the machine. As long as the models faithfully reflect the machine configurations at every program point in an actual run, the model transformer semantics is said to be "sound". This has essentially the same meaning as the soundness of a logic.

A crucial feature of the MTS method is that it has access to a larger solution space—it can generate correct code which graph coloring and linear scan will never consider, thus opens the door to further optimizations. This improvement is mainly due to the use of static cache replacement. In fact, graph coloring and linear scan are making overly restrictive assumptions about the generated code, thus limiting their solution space. One example is that they assume that once a variable is allocated to a register, it stays in that register for its whole life. But actually a variable may stay in a register for a while, then be saved to the stack, and later be loaded back into a (possibly different) register. Thus the assumptions made by graph coloring and linear scan unnecessarily limit the possible forms of generated code. Also, they prevent an important optimization, *live range splitting*, from being naturally included into the register allocator. As we shall see in this article, MTS is based on fewer assumptions, and can thus generate more flexible code. Moreover, it includes live range splitting in the same pass as register allocation in a seamless fashion. This not only makes the compiler simple, but also makes live range splitting "online". In contrast, most of the live range splitting extensions to graph coloring and linear scan are "offline"—done in a separate pass.

# 2 The Big Picture

The high-level theme of the MTS method is to treat the register file as a "cache" for the stack. The traffic between registers and memory is thus analogous to cache replacement. Both the registers and stack locations may contain variables, but access to the registers is more efficient and is often required by the instruction set architecture. So the goal is to keep variables in registers as long as possible, place them in the stack when necessary, and try to minimize register-memory traffic.

Meanwhile, the static information contained in the models is used to generate instructions to move variables into and out of registers. Bacause these instructions are computed at compile time, this is called *static cache replacement*, as opposed to dynamic cache replacement that is used by the processor's memory cache.

## 2.1 The Input Language

The input language (named UIL) is very simple but expressive enough for most imperative programming constructs.

$$
\begin{aligned}
\text{Program} &\rightarrow \textbf{letrec}\,(\text{Definition}\,*)\,\text{Statement}\,* \\
\text{Definition} &\rightarrow f\colon \lambda xyz.\,\text{Statement}\,* \\
\text{Statement} &\rightarrow s_1;\text{Statement}\,* \\
&\mid x \leftarrow y \\
&\mid x \leftarrow y + z \\
&\mid x \leftarrow [y + z] \\
&\mid [x + y] \leftarrow z \\
&\mid \textbf{if}\ t\ \textbf{then}\ \text{Statement}\,*\ \textbf{else}\ \text{Statement}\,* \\
&\mid f(x, y, z) \\
&\mid v
\end{aligned}
$$

A program consists of a list of definitions and a list of statements. Definitions are procedures which may take zero or more arguments and return a value. A statement may be an assignment, an arithmetic operation, a memory read, a memory write, a branching statement, or a procedure call. Note that this language is basically the first-order language one would get from a functional language after a closure conversion and a CPS (or A-normal form) transformation, where no nested expressions and definitions are involved.

## 2.2 Two Pass Register Allocation

The register allocator contains only two passes, a backward liveness analysis which uncovers live range endings, and a forward abstract interpretation pass which performs model transformations and static cache replacement at the same time, and rewrites the code into its final register form.

Unlike liveness analysis for graph coloring or linear scan, we do not need to build an inference graph, thus the liveness analysis pass only has to mark the endings of the live ranges. An example output from the liveness analysis looks like:

$$
\begin{aligned}
x &\leftarrow 0, \{\} \\
y &\leftarrow x + 1, \{\} \\
z &\leftarrow y + 2, \{y\} \\
&f x z, \{f, x, z\}
\end{aligned}
$$

For each statement, the variables whose live range ends are included in the ending set. For example, the ending set of $z \leftarrow y + 2$ is $\{y\}$, because $y$ will no longer be used after this statement. The ending set ressembles the ending scoping delimiter such as "(let ...)" in a functional language. It enables the abstract interpreter to remove the variable bindings from the environment.

## 2.3 Names

The whole allocation process can be thought of as a game of allocating registers to names in the intermediate language program. This is because names and registers are essentially the same: they bind to values and carry them at runtime. But because we have limited number of registers, we must reuse them if we are going to "simulate" the names with registers.

MTS does not require the input to be in the SSA form, because the allocation pass will implicitly perform single assignment renaming: when a variable is assigned, its old binding in the allocation will be removed, and new binding created. The following statements will then be rewritten using the new binding, effectively does a renaming.

Also unlike the SSA form, the same name can be assigned in both branches of a branching statement, for example,

$$\begin{aligned}
&\text{if } x > 0 \\
&\quad \text{then } \quad y \leftarrow 1 \\
&\quad \quad \quad \quad \quad z \leftarrow 2 \\
&\quad \text{else} \\
&\quad \quad \quad \quad \quad z \leftarrow 1 \\
&\quad \quad \quad \quad \quad y \leftarrow 2
\end{aligned}$$

If the two branches result in different allocations, a sequence of shuffling instructions must be inserted under one of the branches in order to bring the two different allocations together. Although the shuffling code can be short, we may want to eliminate it by passing a map of "preferred allocations" for variables already allocated from one branch, so that we can try allocating them to the same registers in another branch.

# 3 Model

The general defintion of a model is an abstract representation of the real world. Usually, a model only represents a few aspects of the real world, but faithfully. As an alternative point of view, a type in a modern type system can be thought of as a model—a model which reflects the possible domain of values of an expression.

In the context of this article, a model is an abstraction of the physical processor, a data structure whose contents reflects the actual machine configuration when the code executes on a physical processor. An example of such a model looks like

$$\{x\!:\!r1, y\!:\!r2\}\{z\!:\!\mathrm{fv}_0\}.$$

This is essentially an environment in an abstract interpreter where each variable has a *binding* to a location. This particular model says that variable $x$ is located in register $r1$, variable $y$ is located in register $r2$, and variable $z$ is located in stack location fv0. (We use the convention $\text{fv}_i$ to denote the *stack location* which is at offset $i$ from the stack pointer.) This is basically a map between variables and locations, but notice that this map is split into two parts that models the register file and the stack separately. This splitting is important because a variable may be located both in a register and a stack location at the same time. For example, $x$ can live in both register $r1$ and stack location $\text{fv}_0$, as in the following model.

$$\{x\colon r1, y\colon r2\}\{x\colon \text{fv}_0, z\colon \text{fv}_1\}$$

The purpose of having the variables multi-homing is that we can optimize away some saves when the variable is already saved in the stack.

A model is currently represented in the prototype compiler as an association list, but other pure mapping data structure such balanced trees can also be used for efficiency purposes. Note that a model is a shared structure and should not be destructively modified.

## 3.1 Initial Model

The allocation process is on a procedure-by-procedure basis, so one may wonder how we get an initial model to start the allocation process. The answer is calling conventions.

There are many kinds of calling conventions for various languages. Here we discuss just the one implemented in the prototype compiler, which turns out to be quite general. Basically, we can think of each UIL procedure as a procedure that takes one additional argument, the return address. We will see that this high-level view simplifies the problem of setting up calls.

The initial model tells us where the arguments (including the return address) are located. It is generated on a procedure-by-procedure basis by a helper function. First of all, we put as many arguments as possible into registers. The rest go into stack locations. There is a special name in the initial model, RET, which denotes the return address. Initially it is bound to the return address register, but may be rebound later. An example initial model looks like

$$\{x\colon r1, y\colon r2, z\colon \text{fv}_0, \mathbf{RET}\colon r0\},$$

where $r0$ is assumed to be a dedicated return address register.

An advantage of treating the return address uniformly with other arguments is that the register used by the return address may be evacuated for other purposes when we run out of registers. An alternative would be to save the return address into the stack, and then restore it to the return address register just before the procedure returns. But we are wasting our time saving and restore the return address if we have enough registers in the procedure. Treating the return address as a normal variable can naturally eliminate this overhead when possible.

# 4 Model Transformers

Model transformers relate models to each other. They may take arguments in addition to the model itself. The arguments may contain directions as how to transform the model. All model transformers are of similar types:

$$(\text{Model}, \text{ExtraArgs}, \ldots) \to (\text{Model}, [\text{Inst}], \ldots).$$

That is, a transformer takes as input a model and some extra arguments, and it produces a new model and a list of necessary instructions for transforming the model.

The list of instructions may contain arbitrary machine instructions such as SAVE, LOAD or MOVE, which can essentially perform live range splitting at appropriet places, as we shall see later on.

## 4.1 Primitive Model Transformers

There are two kinds of model transformers, primitive transoformers and compound transformers. Primitive transformers are basically helpers for the compound transformers. The compound transformers dispatch on the syntax of the intermediate language. There are only three primitive model transformers: `save`, `load`, and `shuffle`.

$$\begin{aligned}
\text{save} &:: (\text{Model}, [\text{Var}]) \to (\text{Model}, [\text{Inst}]) \\
\text{load} &:: (\text{Model}, [\text{Var}]) \to (\text{Model}, [\text{Inst}]) \\
\text{shuffle} &:: (\text{Model}, [(\text{Loc}, \text{Loc})]) \to (\text{Model}, [\text{Inst}])
\end{aligned}$$

The similarity in these three operations is that they all do something "in parallel" on multiple elements (variables or locations), so care must be taken when there are data dependencies between the source and the destination. The functionalities of the first two primitives are almost obvious, but they also contain optimizations for reducing memory traffic.

- `save`. Save a list of variables into stack. Try to find each variable in the stack first. If the variable is already in the stack, skip the saving (emitting no instructions) and return the model intact. Otherwise, find an unoccupied stack location and bind the variable to that location. Leave the orginal binding of the variable with the register alone. After saving, the variable lives in both the register and the stack.

- `load`. Load a list of variables into registers. If a variable is in a register, do nothing but passing on the model. Otherwise the variable is in stack. Emit instructions to load the variable into an unoccupied register, return the updated model which binds the variable to both a register and a stack location. If we run out of registers, some variable must be chosen as an victim to be evicted from registers. We save the victim to a stack location and load the new variable into the victim's home. Note that care must be taken when loading multiple variables in order to prevent the variables in the list to evict each other.

  The way we choose victims is a crucial decision which has impact on the register-memory traffic. This is similar to the different cache replacing policies in the processor's memory cache. We delay the discussion of this topic to a later section.

- `shuffle`. Take a mapping of locations, for example $\{r0 \to r1, r1 \to r2, r2 \to r0, r3 \to r4, r4 \to r5\}$, and generate a list of instructions that accomplishes the moves simultaneously. There may be data dependencies between the locations that need shuffling, the example above shows such a situation where the locations form a loop ($r0 \to r1 \to r2 \to r0$). First of all, the moves are broken down into either paths or loops, then code for appropriete categories are generated. For example, the above mapping contains a loop ($r0 \to r1 \to r2 \to r0$) and a path ($r3 \to r4 \to r5$). The number of instructions generated for a loop shuffle is $n + l$, where $n$ is the number of variables involved, and $l$ is the number of loops. This additional $l$ is because we need a temporary location and an extra instruction to store the entry point value of each loop.

## 4.2 Compound Model Transformers

The compound model transformers dispatch on the syntax of the intermediate language. They are basically the branches of an abstract interpreter which analyzes and transforms the input code. Each branch of it is of the type

$$(\text{Exp}, \text{Model}, \text{Ctx}) \to ([\text{Inst}], \text{Value}, \text{Model}).$$

The input type signifies that the transformer dispatches on `Exp`, the type of the syntax tree node of the intermediate language. It takes a model and a context (`Ctx`), and returns a list of instructions, a value, and a new model. This is basically an effect-value system, where the instruction list ([Inst]) contains the *effects* and the value contains a pure *value*. The effect can be combined with a sequence of other effects, but only the value can be plugged into a context where a value is expected (for example, an operand of a binary operation). The `Ctx` parameter is a "context type" specifying the outer context (e.g., tail context). The model transformers can then look at the context type and decide whether to generate code for a tail call or a non-tail call.

Because the complexity of the compound transformers, they will not be displayed here. The interested reader is referred to the actual code for more information. Here we just briefly discuss the cases one by one.

- $x \leftarrow y$ {end*}. The assignment. First, recursively allocate registers for $y$ and get a new model $model_1$. Notice that $y$ here can match either a variable, a value or a binary operation such as $a + b$. After that, delete end* from $model_1$, use this model to allocate a register to $x$, and get $model_2$. The generated instructions are thus concatenated and returned, together with the rewritten assignment and $model_2$.
- If $t$ then $c$ else $a$. Allocate registers for t, get $model_1$. Then use $model_1$ to allocate for $c$ and $a$, get $model_2$ and $model_3$, respectively. Then generate the "shuffle instructions" that unites $model_2$ into $model_3$, resulting in $model_4$. Finally, combine the code from the recursive call, insert shuffle instructions into the "then" branch. Output the rewritten if-statement and $model_4$.
- $f x y z..., \{\text{end*}\}$. The case for a call is more involved.
    - First, a list of call-live variables are found by deleting end* from the input model. If we are in tail context, there will be no call-lives.

- ○ Then we generate a new map, $\text{map}_1$ by binding the call-lives to their destinations. In short, $\text{map}_1$ specifies the moves we need to make in order to set up the call.
- ○ A frame pointer (stack pointer) adjustment is computed from the number of call-live variables.
- ○ The list of parameters are bound according to the calling convention in $\text{map}_1$.
- ○ If we are in tail context, the special variable `RET` is bound to the return address register in $\text{map}_1$. Otherwise we generate a new label $\text{lab}_1$, and have it bound to the return address register in $\text{map}_1$.
- ○ Shuffling code is generated from $\text{map}_1$.
- ○ The call is rewritten as a list of shuffling code, an adjustment to the frame pointer, a jump, and a reverse adjustment to the frame pointer.
- $s_1, s_2, \ldots$. The sequence. First, the head of the sequence is recursively allocated by passing `nontail` as the `Ctx` parameter, getting $\text{model}_1$. Then using $\text{model}_1$, the rest of the sequence is recursively allocated by passing the input Ctx parameter as the context.
- $[x + y] \leftarrow z$. The memory mutation operation. First we use the primitive model transformer `load` to allocate for $x$, $y$ and $z$, getting $\text{model}_1$. Then we rewrite the statement using $\text{model}_1$.
- $x + y$. Binary operations are similar to memory mutation, thus omitted here.

## 5   Online Live Range Splitting

Live range splitting can be found in several extensions to graph coloring and linear scan. Most of them require a separate pass to split the live ranges, and use the split live ranges to compute the register allocations. MTS does live range splitting in a seamless and implicit manner. You don't see explicit code for it, but it is implicit in the generated instructions. A simple example should clarify this point.

$$\begin{aligned} x &\leftarrow 1, \{\} \\ y &\leftarrow 2, \{\} \\ z &\leftarrow y + 1, \{\} \\ &\vdots \\ w &\leftarrow x + y, \{x\} \end{aligned}$$

**Figure 1.**  Example program with two registers.

In this example, we have only two registers $r1$ and $r2$. After $x$ and $y$ taking $r1$ and $r2$, we run out of registers when we try to find a register for $z$. A graph coloring algorithm would report a failure here unless a live range splitting pass first segments the live ranges of the variables. It is easy to see that the interference graph would contain a strongly connected component here, making the graph uncolorable.

It is interesting to see how MTS splits the live ranges in this situation. It just swaps $x$ out into a stack location, say $fv_2$, then takes its register as the home for $z$. The generated code would look like

$$\begin{aligned}
r1 &\leftarrow 1 \\
r2 &\leftarrow 2 \\
\color{red}{fv_2} &\color{red}{\leftarrow r1} \text{ ; store} \\
r1 &\leftarrow r2 + 1 \\
&\vdots \\
\color{green}{r1} &\color{green}{\leftarrow fv_2} \text{ ; load} \\
r1 &\leftarrow r1 + r2
\end{aligned}$$

**Figure 2.** Generated code containing splitted ranges.

The instruction $fv_2 \leftarrow r1$ is generated in such a manner that as if $x$ goes out of range at that point. After that, $r1$ starts to represent $z$, and reverts back to $x$ after we load $fv_2$ into $r1$ again. Notice that we load $x$ back into $r1$ just by coincidence here. It may be loaded into any other unoccupied register.

If we were to do live range splitting first, we would have to split the live range of $x$ into three segments, at points delimited by the store and load instructions. After that, the second segment would be "spilled" by graph coloring or linear scan, i.e., the variable $x$ would be in the stack within this range.

## 6 Cache Replacement Policies

At program points where we run out of registers, we must choose a "victim" to be placed into the stack. The question is which variable to pick. This is an important question, because it affects the number of register-memory moves. For example, a particularly bad strategy is "last in, first out" (LIFO): swapping out a variable which just recently came into its live range. This strategy is bad because a recent variable is likely be used very soon. Doing so would cause it to be loaded into a register again, possibly causing another variable to be swapped out.

An arguably good strategy is to pick a variable that *will not be used soon*. This can be implemented as follows. When doing the liveness analysis, we record where each variable is referenced. The register allocation pass can then know the NEXT-USE-POS of each variable at each program point. When we need to pick a victim, we just choose the one with highest NEXT-USE-POS.

Now we argue that this strategy generates the least register-memory traffic. Suppose we did not pick $x$, the variable with the highest NEXT-USE-POS. We picked another variable $y$. Then $y$ will be referenced earlier than $x$, because its NEXT-USE-POS must be less than $x$'s. So when $y$ is referenced, it must be loaded into a register. Because we are likely to be under register pressure, we have to swap another variable (possibly $x$) into stack. The question is, why don't we swap $x$ into stack in the first place? By doing so, we delay the need for a load at our best.

# 7  Related Work

Related work has been mostly graph coloring and linear scan and their extensions, with a few isolated works.

Chaitin [2] proposed formulating register allocation as graph coloring. As we have discussed in the introduction, this formulation puts restrictions on the generated code, thus limiting the possibilities of reducing register-memory traffic.

Linear scan [5] has been proposed as an alternative to graph coloring, but in essence, it can also be viewed as a greedy algorithm for graph coloring: coloring the graph vertices in the order of the program. Thus it suffers from almost the same constraints.

Agat [1] proposed a type theory for register allocation, but the main focus of this work is on purely functional languages. It uses a similar method as model transformation semantics (because it is a type theory), but it doesn't aim at reducing register-memory traffic, thus many optimizations are omitted. For example, it doesn't try to efficiently reuse stack locations at all.

Pereira and Palsberg [6] proposed register allocation by puzzle solving. Although their result is comparable with other state-of-the-art algorithms, their method does not do live range splitting, an important way to open up the door to more optimizations.

None of these works do live-range splitting by default. They all need extensions that does the splitting in a separate pass. For example, the live range splitting extension to linear scan is discussed in [7].

# 8  Implications to Verified Compilers

In the context of verified compiler construction, a register allocation pass is usually not deductively verified, as is the case for other parts of the compiler. This is largely due to the complexity involved in the math of the graph coloring algorithm. For exmaple, the certified C compiler CompCert [4] uses a "validation pass" after each compilation in order to check the consistency of the graph coloring. This essentially is a dynamic check in the compiler's run time (also called "compile time"). The validation must be run after each compilation of the user program, instead of just running once when the compiler is constructed.

Actually most parts of a verified compiler can be verified by a logic system, because they are semantics-based. For example, the CPS transformation and the data representation are rather straightforward to verify by "statically analyzing" the compiler. Static analysis is easy when the program fits into an abstract interpretation framework, but rather hard if it is derived from another branch of mathematics, such as graph theory or group theory.

The MTS method is largely semantics-based—it operates as an abstract interpreter. Thus it is possible that it will lead to a formally *verified* register allocation pass, where it is certified once and for all when the compiler is constructed.

# 9 Future Directions

Currently the register allocation assumes a RISC-like architecture. All operands must be loaded into registers before they are operated on. A possibly fruitful future work is to extend this method to architectures that allow and encourage memory operands and irregular instruction formats.

The current MTS does not do backtracking, thus shuffling code length may not be optimal. Limited backtracking may be added to the method, leading to further reduction to register-memory traffic.

Also the current implementation compacts the stack before each call. This will induce quite some memory-memory traffic because the call-live variables may not be saved into the same stack locations across calls. An optimization would be to pre-assign stack locations to call-live variables, so that their locations are fixed. Once a variable is saved, it will not be saved again or moved to another stack location. A backward scan is necessary to collect the call-live information.

Also the current work lacks experimental results. Comparing by using benchmarks is the next topic on the agenda.

# 10 Conclusion

This article has presented a semantics-based method (MTS) for register allocation. The main ideas are model transformer semantics and static cache replacement. Relationships with graph coloring and linear scan are made clear by careful reasoning about their behavior. MTS has been shown to have the advantages of exploring a larger solution space than naive graph coloring and linear scan, and thus possibly leads to optimizations by the other two methods. Live range splitting, an important optimization which is found as an extension to graph coloring and linear scan, is included into MTS seamlessly, without increasing the complexity of the implementation. The semantics-based approach also opens the door to the simplifications of a verified register allocation pass.

**Acknowledgements**
   The authors would like to thank Andy Keep and Oleg Kiselyov for their insightful discussions.